\documentclass[preprint]{aastex62}
\usepackage{tikz}
\usetikzlibrary{decorations.pathmorphing}
\usepackage{tikz}
\usepackage{graphicx}
\usepackage{multirow}
\usepackage{amsmath}
\usepackage{threeparttable}
\usepackage{float} 
\received{}
\revised{}
\accepted{}
\submitjournal{ApJ}

\shorttitle{the soft GeV Spectrum of the Extended Radio Nebula}
\shortauthors{Bao \& Chen}

\begin{document}
\title{\Large\textbf{On the Gamma-ray Nebula of Vela Pulsar -II. the Soft Spectrum of the Extended Radio Nebula}}

\correspondingauthor{Yang Chen}
\email{ygchen@nju.edu.cn}

\author{Yiwei Bao}
\affil{Department of Astronomy, Nanjing University, 163 Xianlin Avenue, Nanjing 210023, China}

\author{Yang Chen}
\affil{Department of Astronomy, Nanjing University, 163 Xianlin Avenue, Nanjing 210023, China}
\affiliation{Key Laboratory of Modern Astronomy and Astrophysics, Nanjing University, Ministry of Education, Nanjing, China}




\begin{abstract}
The Vela\,X pulsar wind nebula (PWN) is characterized by the extended radio nebula (ERN) and the central X-ray ``cocoon". We have interpreted the $\gamma$-ray spectral properties of the cocoon in the sibling paper (Bao et al.\,2019); here, we account for the broadband photon spectrum of the ERN. {Since} the diffusive escape of the electrons from the TeV emitting region is expected to play an insignificant role in shaping the spectrum of the ERN, we attribute the GeV cutoff of the ERN to the reverse shock-PWN interaction. Due to the disruption of the reverse shock, most of plasma {of the PWN} is driven into the ERN. During the subsequent reverberation phase, the {ERN} could be compressed by a large factor in radius, and the magnetic field in the {ERN} is thus significantly enhanced, {burning off the high energy electrons}. We thus obtain the electron spectrum of the ERN and the broadband spectrum of the ERN are explained satisfactorily.
\end{abstract}


\section{INTRODUCTION}\label{sec:intr}
Thanks to its proximity \citep[at a distance $d=287$ pc according to the VLBI parallax measurement,][]{2003ApJ...596.1137D}, the Vela pulsar and its associated pulsar wind nebula (PWN) Vela\,X can be studied in some detail. The Vela pulsar, with a present spin-down luminosity $\dot E_\textup{now}=7\times 10^{36}$ erg s$^{-1}$ and characteristic age $\tau_c = 11400$ yr \citep{2005AJ....129.1993M}, is located inside the Vela supernova remnant (SNR). The pulsar powers an extended radio nebula (ERN) of size $\sim 1.2^\circ \times 3^\circ$ \citep{1997ApJ...475..224F} and an X-ray ``cocoon'' of $\sim 0.8^\circ$ in length \citep{1997ApJ...480L..13M}. The complicated morphology of the PWN is suggested to arise from the interaction of an asymmetric reverse shock \citep[][{hereafter S18}]{2001ApJ...563..806B,2018ApJ...865...86S}, which originates in the density anisotropy surrounding the Vela SNR. The estimated density is {suggested to be} higher in the northeast and lower in the southwest (S18). H.E.S.S observation detected the TeV $\gamma$-ray counterpart of the cocoon with a hard spectral index $\sim 1.45$ \citep{2006A&A...448L..43A}. The magnetic field strength in the X-ray cocoon is estimated to be $\sim 4$ $\mu$G based on the X-ray to TeV flux ratio \citep{2006A&A...448L..43A}. Fermi-LAT off-pulse observations detected an extended GeV $\gamma$-ray emitting region with a soft spectral index $\sim 2.4$, which is spatially coincident with the ERN \citep{2010ApJ...713..146A}.

{In order} to explain the puzzling two-peaked $\gamma$-ray spectral energy distribution (SED) of the whole {PWN}, {\citet{2008ApJ...689L.125D} proposed that} the spin-down power of the Vela pulsar is converted into two components contributing to the radio and X-ray radiation, respectively. The age of the cocoon is assumed to be $\sim 11$ kyr (the age of the Vela pulsar) in their model. However, the scenario is challenged by the TeV emission beyond the cocoon as detected by H.E.S.S \citep{2012A&A...548A..38A}. The TeV emission is extended over the ERN, indicating that the cocoon and the ERN are mutually associated. \citet{2011ApJ...743L...7H} argue that diffusive escape of electrons should be introduced to solve the dilemma, and the required diffusion coefficient is $D_\textup{ERN}=10^{26}(E$/100\,GeV)\,cm$^2$\,s$^{-1}$, where $E$ is the energy of the electrons. The cocoon in their model is assumed to be formed 230 years ago, and the soft GeV spectrum of the ERN stems from diffusive escape of TeV electrons. Nevertheless, recent observations {seems to indicate a very slow diffusion. Radio observation has revealed that the magnetic field in the ERN is $B_\textup{ERN}=$10--50 $\mu$G based on the width of the filaments \citep{1995MNRAS.277.1435M}. This field, however, would be even more stronger when the ERN is compressed by the reverse shock. The diffusion coefficient is estimated to be $\approx 1\times 10^{26}(E$/10 TeV)$^{1/3}$\,cm$^2$\,s$^{-1}$ in the TeV nebula \citep[][hereafter, Paper I]{paperI}, in which the magnetic field is estimated to be $4\,\mu$G in strength \citep{2006A&A...448L..43A}. Since the diffusion coefficient is expected to be inversely proportional to the magnetic field, the diffusion coefficient in the ERN should be $D_\textup{ERN} \sim 3 \times 10^{24}(E/100$\,GeV)$^{1/3}$($B$/30$\,\mu$G)$^{-1}$\,cm$^2$\,s$^{-1}$, which is much smaller than that is required to diffuse enough TeV electrons out of the ERN}.

{While Paper I is devoted to the diffusion of the electrons of the TeV nebula, in this paper, we aim to develop new scenario to account for the broadband emission of the Vela\,X PWN by developing a new scenario based on the hydrodynamic simulation in S18}. {The ERN contains most of the plasma \citep[with an energy $\sim 5\times 10^{48}$ erg,][]{2010ApJ...713..146A} injected before the reverse shock-PWN interaction.} Because the diffusion coefficient in the ERN {is expected to be small}, diffusive escape {may be neglectable} in the ERN. Instead, the reverberation phase experienced in the earlier evolution of the PWN, although short, is suggested to play a significant role in the energy loss of the electrons of the PWN. In this phase, the PWN is compressed by the reverse shock that moves inwards, reverberating several times and oscillating \citep{2006ARA&A..44...17G}. The PWN could be compressed by a large factor in radius, and the magnetic field strength in the PWN is accordingly increased \citep{2006ARA&A..44...17G}. The enhanced magnetic field gave rise to a high X-ray luminosity (which might be even higher than the spin-down luminosity of the pulsar, \citeauthor[see, e.g.,][]{2018ApJ...864L...2T}\,2018) and the highest energy electrons were thus burnt off (\citeauthor{2006ARA&A..44...17G}\,2006; the synchrotron lifetime for electrons with energy 700 GeV is {$t_\textup{sync}=4.4 \times 10^3(B/60\ \mu\textup{G})^{-2}$ yr}, \citeauthor{2011ApJ...743L...7H}\,2011), and therefore the soft GeV spectrum of the ERN can be explained naturally. Our model is described in \S 2, model calculation is presented in \S 3, and the discussion and conclusion are given in \S 4.  

\section{MODEL DESCRIPTION}
In paper I, we approximate that the TeV nebula contains a small fraction of plasma injected impulsively upon the formation of the cocoon at $\tau_\textup{s}$ {($\sim$ 8\,kyr, the time when the SNR reverse shock squeezes the ERN out of the original PWN, S18).} The cocoon is not compressed by the reverse shock, and therefore the TeV electrons survive. In this paper we lay stress on the SED of the ERN. According to the hydrodynamical simulation in S18, the reverse shock contacts the PWN in the northwest at an age of $\sim 4$ kyr. At $\tau_\textup{s}$, the reverse shock coming from the northeast sweeps over the whole PWN, disrupting it, {driving most of plasma into the ERN}, and creating a tail (which becomes the X-ray cocoon later) to the south. {At $T_\textup{age}$ ($\sim 12$ kyr, the total age of the PWN), the cocoon and the ERN are separated, which accounts for the spatial distribution of GeV and TeV emission.} As can be seen from S18, spherical symmetry appears to be a good approximation in the first stage (0--4 kyr). Before the reverse shock-PWN interaction, the unshocked ejecta is cold, and a shock forms as the PWN expands supersonically. Following \citet{1984ApJ...280..797C}, we assume the PWN is surrounded by a thin shell of swept-up supernova ejecta, and the whole PWN expands isobarically with the expanding thin shell, and the expansion of the PWN can be calculated numerically. The expansion of PWN {before the reverse shock-PWN interaction} can be described by

\begin{eqnarray}
\label{expansion}
\frac{d R(t)}{dt}=v(t),\\
M_\textup{sh}(t)\frac{d v(t)}{dt}=4 \pi R^2(t) \left[P_\textup{PWN}(t)-P_\textup{ej}(R,t)-\rho_\textup{ej}(v-v_\textup{ej})^2 \right],\\ \frac{d M_\textup{sh}(t)}{dt}=\left \{
\begin{array}{ll}
4 \pi R^2(t) \rho_\textup{ej}(R,t) (v(t)-v_\textup{ej}(R,t)) &\ \ \ v_\textup{ej}(R,t)<v(t),\\
0 &\ \ \ v_\textup{ej}(R,t)>v(t),
\end{array} \right. 
\end{eqnarray}
where $P_\textup{ej}(t)$, $\rho_\textup{ej}(t)$, and $v_\textup{ej}$ represent the pressure, density, and velocity of the SNR ejecta {\citep[obtained following][]{1999ApJS..120..299T}}, respectively; $M_\textup{sh}(t)$, $v(t)$, and $R(t)$ represent the mass, velocity, and radius of the thin shell, respectively; and $P_\textup{PWN}(t)$ represents the isobaric pressure of PWN. When the PWN {encounters} the reverse shock, the ejecta ahead of the shell has been heated by the reverse shock, the forward shock of the swept ejecta shell no longer exists, and hence the thin shell approximation no longer holds. The radius evolution in the subsequent reverberation phase {($\sim$4--12 kyr) is thus much more complicated. Meanwhile, the asymmetry perplexes the evolution further. Fortunately, some qualitative inferences can be drawn to constrain the evolution of the radius: (a) Since the cocoon arises from compression (S18), the original PWN is expected to be shrinking at $\tau_\textup{s}$, {i.e., $v(\tau_\textup{s})<0$}; (b) The PWN shrinks at almost constant velocity to about half of its maximum value in the first compression \citep[which lasts thousands of years,][]{2001A&A...380..309V}; and (c) A complete circle of reverberation lasts several thousand years longer than the first compression \citep[][]{2001A&A...380..309V}, and therefore only one re-expansion is expected during $\tau_\textup{s}$--$T_\textup{age}$ ($\sim$8--12 kyr). We thus} assume that the radius of the PWN (the ERN) shrinks at a constant velocity to $R_\textup{min}$ at an age of $T_\textup{min}$, and then the PWN expands at another constant velocity until its radius reaches the present size $R_\textup{now}$.
({More complicated radius evolution may not substantially change the case; see the last paragraph of \S3.}) The relatively faint TeV radiation {and bright X-ray emission} in the immediate vicinity of the Vela pulsar indicates a strong magnetic field is formed after the interaction of the reverse shock \citep[see e.g.,][]{2011ApJ...743L...7H}. {Meanwhile, the TeV flux indicates that the total energy in the TeV nebula is low \citep[$\sim 10^{46}$ erg,][]{2010ApJ...713..146A}, implying that TeV electrons injected after the interaction are cooled due to the magnetic field near the pulsar. Although the ERN has been driven away from the pulsar, the plasma injected from the pulsar can still flow into the ERN through the cocoon (S18). To account for the faint TeV emission, we thus assume {that} after {the passage of the reverse shock ($\tau_\textup{s}$), a strong magnetic field forms in the vicinity of the pulsar, and} the injected electrons are cut off {(see Paper I)} at the Lorentz factor of $\gamma_\textup{cut}$ because of substantial synchrotron loss ($t_\textup{sync}\sim 10^2(B/400\ \mu\textup{G})^{-2}(E/700\ \textup{GeV})^{-1}$ yr).}

The electron spectrum can be obtained by solving the electron number conservation equation:
\begin{equation}%
\label{diffloss}
\frac{\partial N(\gamma,t)}{\partial t}=\frac{\partial}{\partial \gamma}\left[\dot{\gamma}(\gamma , t)N(\gamma,t) \right]+\left \{ 
\begin{array}{ll}
Q_\textup{pairs}(\gamma,t)\gamma^{-\alpha} & \text{for }t \le \tau_\textup{s},\\
Q_\textup{pairs}(\gamma,t)H(\gamma_\textup{cut}-\gamma)\gamma^{-\alpha} & \text{for }t \geq \tau_\textup{s},\\
\end{array} \right.
\end{equation}
where $N(\gamma,t)$ is the electron distribution function, $Q_\textup{pairs}(\gamma,t)$ is a normalization constant, $\alpha$ represents the index of injected electrons, and $H(x)$ is the Heaviside step function
\begin{equation*}
H(x)=\left \{
\begin{array}{llc}
0 & \text{for }x \textless 0,\\
1 & \text{for }x \geq 0.
\end{array} \right.
\end{equation*}
The total {electron} energy injection into the PWN is at a rate
\begin{equation}
\dot E(t)=\frac{\left(1-\eta\right)\dot E_0}{(1+\frac{t}{\tau_0})^{{(n+1)}/{(n-1)}}}= \int^{\gamma_\textup{max}}_{1} Q_\textup{pairs}(\gamma,t)\gamma^{-\alpha}\,\gamma\,d\gamma,
\end{equation}
where $\dot E_0$, and $n$ are the initial spin-down luminosity, the braking index of the Vela pulsar, respectively, {$\tau_0=2\tau_c/(n-1)-T_\textup{age}$ is the initial spin-down age of the pulsar, and} {$\eta$ is the fraction of the spin-down energy deposited to the magnetic field.} The energy loss term $\dot \gamma$ is determined by synchrotron, inverse Compton {(off CMB, far infrared, and near infrared photons)}, bremsstrahlung, and adiabatic losses. The magnetic field is determined by the magnetic energy injected and the expansion of the PWN \citep{2010ApJ...715.1248T}
\begin{equation}
\label{eq:B}
\frac{d W_B(t)}{dt}=\eta \dot E -\frac{W_B(t)}{R(t)} \frac{d R(t)}{dt},
\end{equation}
where $W_B(t)$ represents the magnetic energy in the PWN.

\section{MODEL CALCULATION}
\begin{center}
	\begin{figure}[H]
		\includegraphics[scale=0.48]{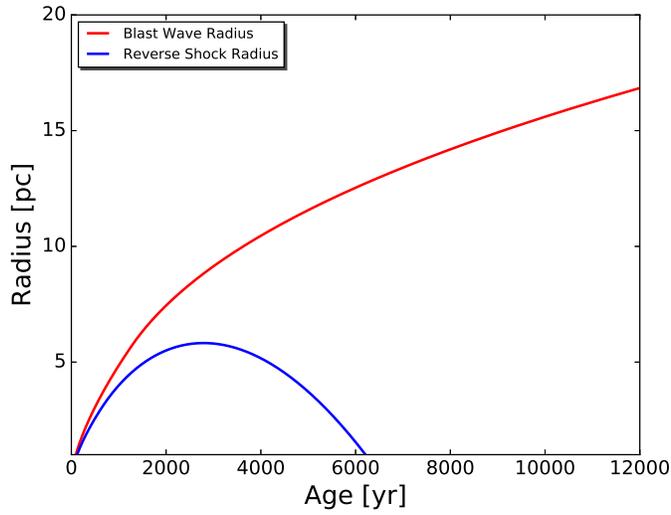}
		\caption{Time evolution of the blast wave and reverse shock of the Vela SNR without the pulsar.}
		\label{fig:SNR}
	\end{figure}
\end{center}

\begin{center}
\begin{figure}[H]
\includegraphics[scale=0.48]{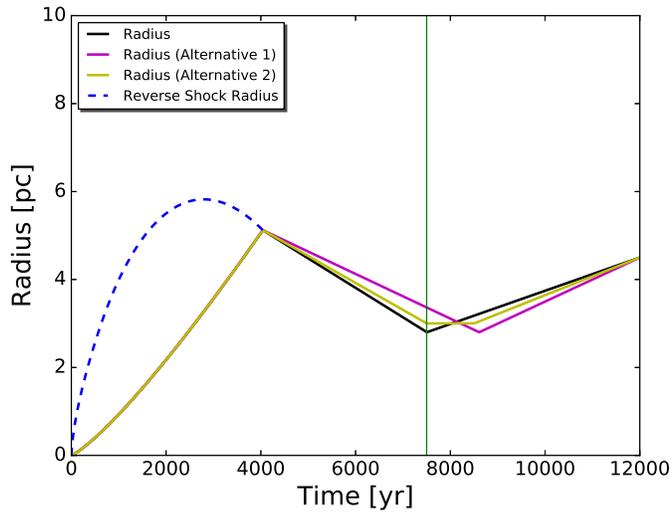}
\includegraphics[scale=0.48]{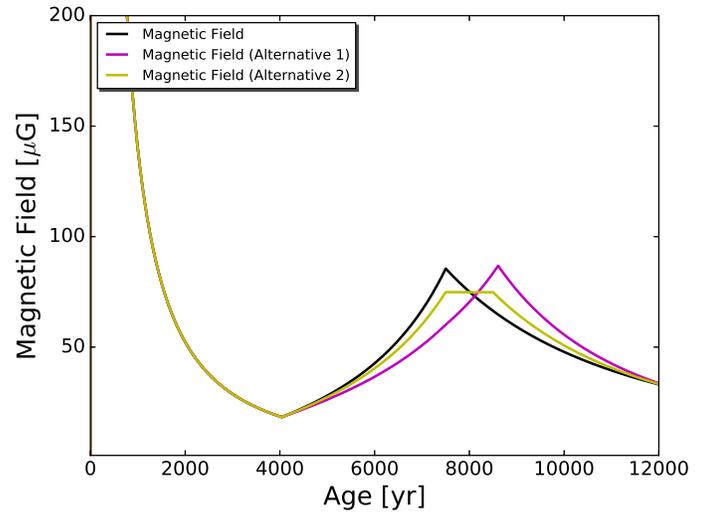}
\caption{{Time evolution of the radius and magnetic field} of the ERN\label{fig:Revo}}
\end{figure}
\end{center}

\begin{center}
\begin{figure}[H]
\includegraphics[scale=0.48]{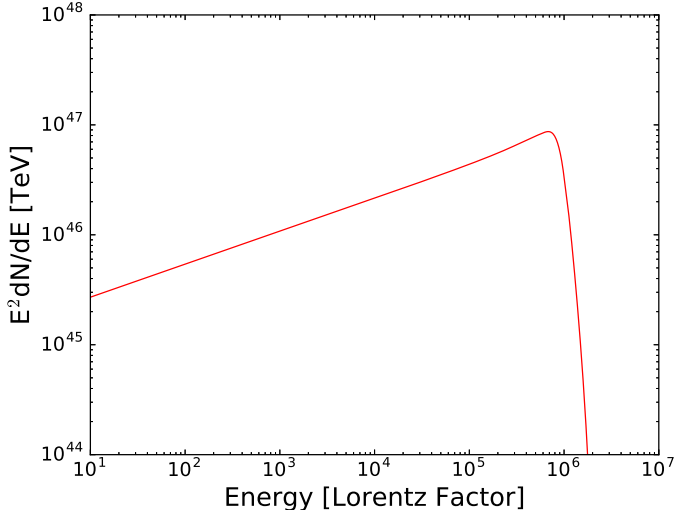}
\caption{The model electron spectrum of the ERN at present}
\label{fig:elec}
\end{figure}
\end{center}

\begin{center}
\begin{figure}[H]
\includegraphics[scale=0.48]{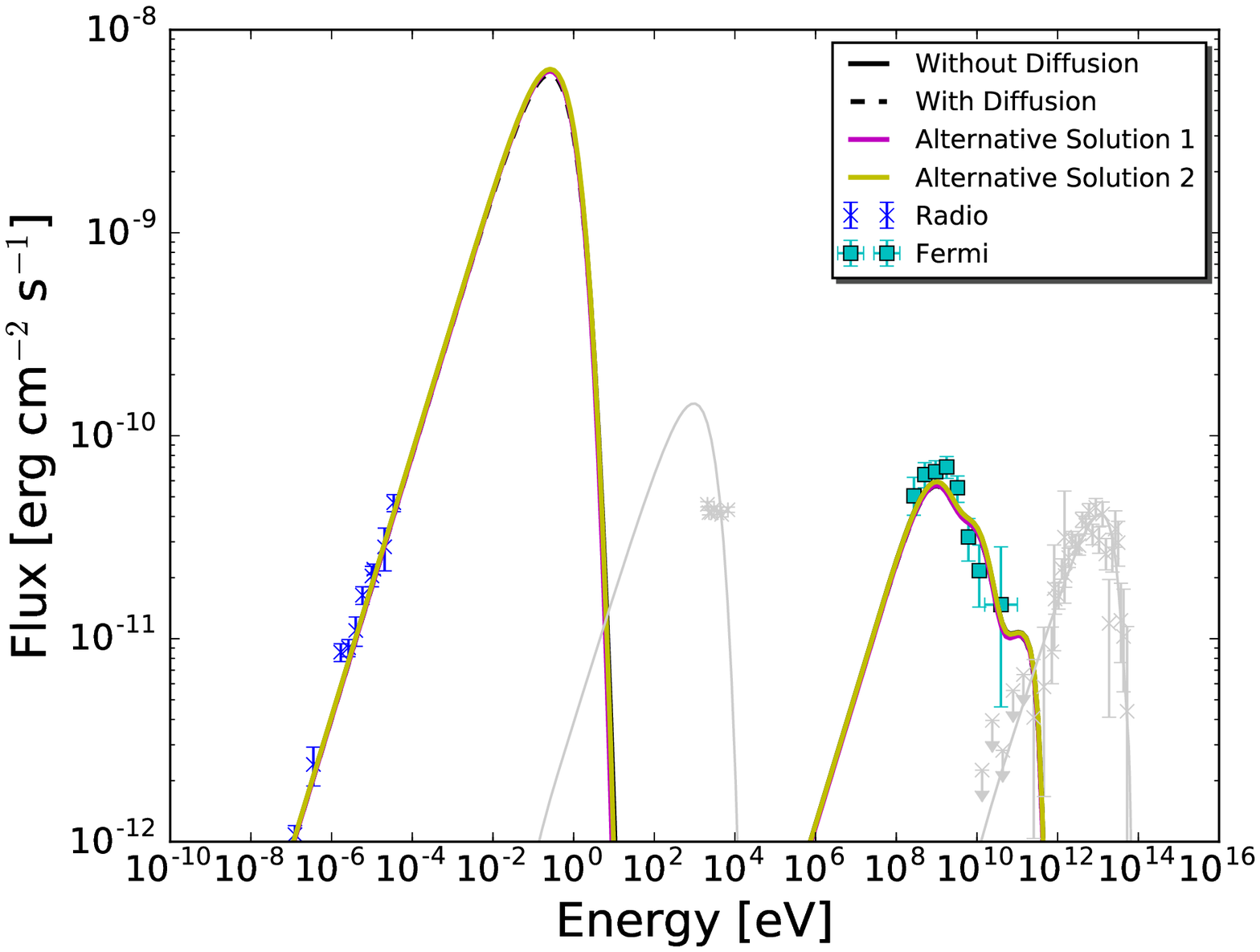}
\caption{The SED of the whole Vela X. The radiation from the ERN is plotted in black, with the radio data taken from {\citet{2001A&A...372..636A}}, the ROSAT X-ray upper limit from \citet{2010ApJ...713..146A}, and the Fermi-LAT data from \citet{2013ApJ...774..110G}. The radiation comes from the cocoon {(calculated according to Paper I)} is plotted in {grey}, with the ASCA X-ray data taken from \citet{2010ApJ...713..146A}, and the $\gamma$-ray data from \citet{2018A&A...617A..78T}. {The Fermi data presented in \citet{2018A&A...617A..78T} are extracted in a small region around the cocoon, and thus do not conflict with the last Fermi data point presented in \citet{2013ApJ...774..110G}.}}
\label{fig:SED1}
\end{figure}
\end{center}

We perform the model calculation for the ERN of the Vela\,X PWN. The braking index of the Vela pulsar is measured to be $n=1.7$ \citep{2017MNRAS.466..147E}. {We first constrain the density of the ambient interstellar medium $n_\textup{ISM}$ and the ejecta mass $M_\textup{ej}$.
We adopt an age of 12 kyr which is similar to the characteristic age of the Vela pulsar.} The angular radius of the SNR is $\sim 3.4^\circ$ \citep{1995Natur.373..587A}, corresponding to $\sim 17$ pc. {Since the energy injected by the pulsar is much smaller than the canonical energy of supernova explosion ($10^{51}$ erg), the existence of pulsar can hardly affect the evolution of SNR blast wave. With the pulsar and its wind nebula neglected, the evolution of the SNR blast wave and the reverse shock are calculated according to \citet{1999ApJS..120..299T}, as shown in \autoref{fig:SNR}. Next, {incorporating} the pulsar and its wind nebula, {we} calculate the radius, magnetic field and electron spectrum evolution of the PWN before the reverse shock-PWN interaction} using Equations \ref{expansion}--3, \autoref{eq:B} and \autoref{diffloss}, respectively. {When the PWN touches the reverse shock at $\sim 4$ kyr (see \autoref{fig:Revo}), the interaction begins, the reverse shock evolution is interrupted by the interaction, and henceforth the blue line in \autoref{fig:SNR} does not hold. Thirdly, using the radius evolution in $\sim 4$--12 kyr as is described in \S 2, we calculate the magnetic field and electron spectrum evolution accordingly. The radius and magnetic field evolution of the PWN are plotted in \autoref{fig:Revo}, and the electron spectrum at present is shown in \autoref{fig:elec}. The magnetic field can be enhanced to $\sim 10^2\,\mu$G during the compression, which burns off the TeV electrons ($\sim 10^3 (B/10^2\ \mu\textup{G})^{-2}(E/700\ \textup{GeV})^{-1}$ yr) injected before $\tau_\textup{s}$}. The SED of the ERN is well fitted (as shown in Figure \ref{fig:SED1}) with the parameters listed in \autoref{tab:par}. Also, in \autoref{fig:SED1}, the SED of the cocoon, which is accounted for in Paper I, is plotted together for comparison\footnote{{The magnetic field strength in the cocoon can be calculated as $B^2=B^2_\textup{eff}-8\pi u_{_{\textup{CMB}}}- 8\pi u_{_{\textup{FIR}}} = $4 $\mu$G, where $u_{_\textup{CMB}}$ and $u_{_\textup{FIR}}$ denote the energy density of the CMB and the FIR photons, respectively; IC off near infrared photons are severely suppressed (Paper I) and thus neglected here.}}. In \autoref{tab:par}, $E_\textup{sn}$ represents the supernova explosion energy (a canonical value $E_\textup{sn}=10^{51}$ erg is adopted), $M_\textup{ej}$ represents the mass of the ejecta; $B_\textup{now}$ represents the resulting magnetic field strength at present; {$T_\textup{NIR}$, $T_\textup{FIR}$, $T_\textup{NIR}$ represent the temperature of the CMB, far infrared and near infrared photons, respectively; $u_{_{\textup{CMB}}}$, $u_{_{\textup{NIR}}}$ represents the energy density of the CMB and near infrared photons, respectively; and $D_\textup{ERN}$ represents the diffusion coefficient in the ERN.}

{The radius evolution of the PWN assumed in \S2 is not unique. More complicated evolution can be compensated by a change of $\sim 7$\% (0.2\,pc) in the value of $R_\textup{min}$. Actually, if we allow alternative radius evolution of the PWN, with the $R_\textup{min}$ changed by $<0.2$\,pc, the modified evolution (shown in \autoref{fig:Revo}) and SEDs (shown in \autoref{fig:SED1}) are very similar to the original ones.}

\section{discussion}
\subsection{{Choice} of the radio data}
{In the model calculation for the SED (\autoref{fig:SED1}), radio fluxes are used and ascribed to synchrotron. Actually, there are two sets of radio data of the ERN in the literature:} the WMAP data presented in \citet{2010ApJ...713..146A} and the radio data presented in \citet{2001A&A...372..636A}. As has been noticed by \citet{2011ApJ...743L...7H}, there is an inconsistency between {them. Here, we have adopted} the data presented in \citet{2001A&A...372..636A} which are more consistent with the single power-law spectrum. More detailed observation should resolve this issue in the future.
	

\subsection{The effect of diffusive escape}
{In the calculation of SED (\S 2), we have neglected diffusive escape of the electrons in the ERN.} {Since} the diffusion coefficient $D \propto 1/B$, the diffusive escape timescale $\tau_\textup{escape}\propto R^2/D \propto BR^2$. {If no energy is injected ($\dot E =0$), it can be obtained from \autoref{eq:B} that $BR^2$ is a constant. Taking energy injection into consideration, $BR^2$ will increase slightly with time.} Meanwhile, the synchrotron lifetime of the electrons is $t_\textup{sync}\propto B^{-2}$. Hence the synchrotron loss overwhelmed diffusive escape {due to the magnetic field strength boost during the compression phase.} {However, even if diffusive escape ({the timescale of which} is adopted to be $\tau_\textup{escape}=R^2/(6D)$ following \citet{1965P&SS...13....9P}) taken into consideration, the SED shows only a slight difference from the former one, as is shown in \autoref{fig:SED1}.}

\subsection{About the density of the ambient medium}
The interstellar medium around the Vela SNR is not uniform. {Using} all the eight values of {the observed ambient medium density} $n_\textup{ISM}$ \citep{1995Natur.373..587A}, we obtain an {average density} $n_\textup{ISM,mean}=0.21$ cm$^{-3}${, and the density $n_\textup{ISM}$ (0.3\,cm$^{-3}$) adopted in fitting is in good accordance with this average value.} The value of $n_\textup{ISM,mean}$ can also be constrained by fitting the present SNR radius: considering the radius evolution presented in \citet{1999ApJS..120..299T}, fixing the age of the SNR $T_\textup{age}=12$ kyr ($\sim \tau_c$), {the SNR radius} $R_\textup{SNR} \sim 17$ pc {is reproduced} when $n_\textup{ISM,mean}=0.3$ cm$^{-3}$ is adopted.


\subsection{{The differences between our model and previous models}}
{The origin of two-peaked $\gamma$-ray spectrum of the Vela X PWN has been debated for a decade. We have (in paper~I and this paper) ascribed the TeV peak to the $\la10^2$\,TeV electrons injected upon the formation of the cocoon and ascribed the GeV peak to the lower-energy electrons in the ERN. While \citet{2008ApJ...689L.125D} suggested that the GeV and TeV $\gamma$-rays arise from two different electron sources, \citet{2011ApJ...743L...7H} attributed the cocoon to the reverse shock-PWN interaction. Although our scenario is similar to that was proposed in \citet{2011ApJ...743L...7H} to some extent, we incorporate following improvements: (a) our physical scenario and spectral calculation essentially step from the simulation presented by S18, which consider the hydrodynamic evolution of PWN, and explain the morphology of the Vela X PWN; (b) we adopt a larger age ($\sim 4$ kyr) of the cocoon according to S18 (while \citeauthor{2011ApJ...743L...7H} adopt an age of a few hundred years) ; (c) we assume a Kolmogorov diffusion which is of weak energy dependence based on the spatial variation of TeV $\gamma$-ray indices presented in \citet{2012A&A...548A..38A} (see Figure 3 in Paper I), while \citeauthor{2011ApJ...743L...7H} assume a Bohm diffusion; (d) we obtain a stronger magnetic field ($33\,\mu$G) in the ERN, which is consistent with the radio observation \citep{1995MNRAS.277.1435M}, while \citeauthor{2011ApJ...743L...7H} assume a weak magnetic field ($4\,\mu$G).}

\section{Summary}
{In terms of an elaborate hydrodynamic simulation, S18 successfully provide an overall interpretation of the evolution and the present morphology of the Vela\,X PWN, which is comprised of a newly formed X-ray cocoon and an relic ERN. This simulation appears to have paved the way to disentangle the puzzling property of the $\gamma$-ray emission. The passage of the SNR reverse shock squeezes the ERN out of the original PWN, while a small fraction of plasma is left uncompressed and forms the cocoon. Paper\,I shows that the diffusion of the TeV electrons of the cocoon gives birth to the TeV nebula. 
In this paper, we have accounted for the broadband photon spectrum of the ERN. We estimate that the diffusive escape of the electrons from the TeV emitting region plays an insignificant role in shaping the spectrum of the ERN. 
		During the reverberation phase of the ERN, {it is} compressed by a large factor in radius, and the magnetic field in the ERN is hence boosted and exhausted the high-energy electrons inside. We thus obtain the electron spectrum of the ERN, and the broadband spectrum of the ERN (especially, with the soft $\gamma$-ray cutoff therein) is explained satisfactorily.}
	
	 PWN has been suggested to be a natural explanation for unidentified $\gamma$-ray source for years \citep{2009arXiv0906.2644D,2018arXiv180511522K}. Once its unusual multiwavelength morphology and $\gamma$-ray spectrum are understood, Vela-like PWNe which lies farther from us can be understood better. {Elaborate MHD simulations incorporating radiative losses may look deeper into the physics in the PWNe such as Vela~X, and future high-energy observation with improved resolution can play an important role in disentangling the spatial and spectral properties of Vela X and Vela-like PWNe.}

\acknowledgments
{We thank the anonymous referee for constructive comments, and Patrick Slane, Rino Bandiera, and Xiangdong Li for helpful discussion}. This work is supported by the 973 Program under grants 2017YFA0402600 and 2015CB857100 and the NSFC under grants 11773014, 11633007, and 11851305.

\begin{center}
\begin{deluxetable}{p{2.5cm}cc}
\tablecaption{Fitting Parameters\label{tab:par}}
\tablewidth{0pt}
\tablehead{
  Parameter     & Quantity 
}
\startdata
$T_\textup{age}$ (yr)                       & 12000 \\
$\tau_c$ (yr)                               & 11400 \\
$n$ 										& 1.7 \\
$M_\textup{ej}$	($M_\odot$)					& 5 \\ 
$E_\textup{sn}$ (erg) 						& $10^{51}$ \\
$n_\textup{ISM}$ (cm$^{-3}$)                & 0.3 \\
$d$ (pc)									& 287 \\
$R_\textup{ERN,now}$ (pc)					& 4.5 \\
$T_\textup{min}$ (yr)						& 7500 \\ 
$R_\textup{min}$ (pc)						& 2.8 \\
$\alpha$                                    & 1.7 \\
$\gamma_\textup{max}$ 						& $2 \times 10^9$ \\
$\gamma_\textup{cut}$						& $1 \times 10^6$ \\
$\eta$ 										& 0.15 \\
$B_\textup{now}$  ($\mu$G)					& 33  \\  
$\tau_\textup{s}$ (yr)						& 7500 \\ 
$T_{_{\textup{FIR}}}$~(K)					& 2.73 \\    
$u_{_{\textup{FIR}}}$~(eV\,cm$^{-3}$)		& 0.25 \\     
$T_{_{\textup{FIR}}}$~(K)					& 25 \\    
$u_{_{\textup{FIR}}}$~(eV\,cm$^{-3}$)		& 0.2 \\        
$T_{_{\textup{NIR}}}$~(K)					& 3000 \\    
$u_{_{\textup{NIR}}}$~(eV\,cm$^{-3}$)		& 0.3 \\  
$D_\textup{ERN}$~(cm$^2$\,s$^{-1}$)		& $10^{26}(E$/10TeV)$^{1/3}\,(B/4\mu \textup{G})^{-1}$\\           
\enddata
\end{deluxetable}
\end{center}

\end{document}